\documentclass[12pt,a4paper]{article}

\usepackage[utf8]{inputenc}
\usepackage[T1]{fontenc}
\usepackage{amsmath,amssymb,amsfonts}
\usepackage{mathtools}
\usepackage{graphicx}
\usepackage[margin=1in]{geometry}
\usepackage{cite}
\usepackage{float}
\usepackage{enumitem}
\usepackage{bm}

\usepackage{xcolor}
\usepackage{setspace}
\usepackage{braket}
\usepackage{hyperref} 

\newcommand{\WT}{\mathcal{W}_T}
\newcommand{\WM}{\mathcal{W}_0}
\newcommand{\Rmn}{\mathcal{R}_{mn}}
\newcommand{\ellmn}{\ell_{mn}}

\newcommand{\Feq}{\dot{F}_{\mathrm{eq}}}
\newcommand{\Finst}{\dot{F}_\tau}
\newcommand{\Corr}{\mathcal{C}}

\newcommand{\bv}{\mathbf{v}}

\newcommand{\arcsinh}{\operatorname{arcsinh}}

\title{%
Unruh–DeWitt Detector Response in Toroidal Spacetime
}

\author{%
{Nirmalya Kajuri \footnote{nirmalya@iitmandi.ac.in} and Sheeshram Siddh \footnote{di2206@students.iitmandi.ac.in}}\\[4pt]
\textit{School of Physical Sciences,}\\[2pt]
\textit{Indian Institute of Technology Mandi,}\\[2pt]
\textit{Himachal Pradesh, India}
}

\date{\today}

\begin{document}

\maketitle

\begin{abstract}
The global topology of spacetime, though invisible to local curvature
measurements, leaves signatures on the correlation functions of quantum
fields. We study these signatures using an Unruh-DeWitt
particle detector operating in four-dimensional Minkowski spacetime
with two spatial directions periodically identified, yielding a
spatial topology $\mathbb{R}\times T^2$. 
We compute detector transition rates for three
trajectories: uniform inertial motion, uniform proper acceleration
directed along one of the compact axes, and uniform proper
acceleration along the non-compact axis. Our results show how a local quantum measurement can reveal features of the large-scale spatial topology.
\end{abstract}

\newpage
\tableofcontents
\newpage

\section{Introduction}
\label{sec:intro}

General relativity determines the local curvature of spacetime through
the Einstein field equations. But the global shape of space is not determined by the field equations. 

In cosmology, topologically nontrivial spatial geometries have been considered as possible models of the universe. For example, a spatially flat cosmological model is equally consistent with infinite Euclidean sections $\mathbb{R}^3$ and with many compact manifolds: flat tori, chimney spaces, and other multiply-connected Euclidean three-manifolds~\cite{Lachieze-Rey:1995qrb, Levin:2001fg}. Discriminating among these possibilities is still a fundamental question at the level of topology and physics~\cite{Cornish:1997ab, Cornish:1997hz, Cornish:1997rp, Luminet:2003dx, Aurich:2004fq, Greene:2025bkp, Greene:2025qwq} and one which is under active observational investigation \cite{COMPACT:2022gbl,COMPACT:2022nsu,COMPACT:2023rkp,COMPACT:2024dqe,COMPACT:2024cud,COMPACT:2024qni,Samandar:2025kuf,COMPACT:2025adc,COMPACT:2026pqs,COMPACT:2026vdj}.

Traditional approaches to probe the cosmic topology are
typically non-local: they look for repeated patterns across the sky, such
as the cosmic microwave background~\cite{Cornish:1997ab, Cornish:1997rp, Mota:2008jr, Bielewicz:2011jz} or crystallographic signatures in
the distribution of distant objects~\cite{Luminet:2003dx}. A natural
question to ask is whether the topology of space leaves some trace that is detectable in measurements that are locally performed by an observer.

A promising approach to this question involves Unruh--DeWitt detectors~\cite{Unruh:1976db, DeWitt:1980hx}: a
two-level quantum system moving along a worldline and
coupled to a scalar field. UdW detectors capture the effects of radiation and vacuum fluctuations in a single measurable quantity: the transition rate. In Minkowski spacetime, a
uniformly accelerated detector registers the Planck thermal
distribution at the Unruh temperature $T = a/(2\pi)$, while an inertial
detector registers nothing~\cite{Unruh:1976db}. For a review of Unruh effect we refer to \cite{Crispino:2007eb}.

The mode expansion along the compact
directions is modified, and the Wightman two-point function acquires a new structure reflecting the global identifications. Because the Unruh–DeWitt transition rate is built directly from this two-point function, any topological modification of the two-point function is immediately registered in the detector’s response. Unruh effect as a probe of nontrivial topology has been studied in~\cite{Langlois:2005if, Lin:2015aua, Chiou:2016exd, Wilkinson:2024rva}.

An especially thorough study of the single compact-dimension case was
carried out in \cite{Chiou:2016exd}, where the Unruh--DeWitt
detector was analyzed on a spacetime of topology
$\mathbb{R}^{2,1}\times S^1$. That work demonstrated that an inertial detector's de-excitation rate picks up a correction
depending on the compactification length and the detector's velocity along the compact direction; that uniform acceleration along the compact direction destroys the stationarity of the trajectory with respect to the vacuum; and that acceleration along a noncompact direction leaves the trajectory stationary.

In this paper we take a natural further step and study the
Unruh-DeWitt detector when \textit{two} spatial dimensions are
simultaneously compactified, producing a spatial topology
$\mathbb{R}\times T^2$. The passage from one compact dimension to two is not a mere
repetition. The image structure of the Wightman function becomes a
two-dimensional lattice rather than a one-dimensional chain, and the
resulting physics is quite different. Transition rates depend on a lattice norm that encodes both compactification lengths and their ratio, while the set of critical times at which the instantaneous transition rate diverges acquires a correspondingly richer structure.

We examine three kinematic configurations in detail. First, uniform
inertial motion, for which the detector is in equilibrium with the
field and the transition rate can be computed by contour integration.
Second, uniform proper acceleration directed along one of the compact
dimensions, for which equilibrium breaks down and only the
instantaneous transition rate has physical meaning. Third, uniform proper acceleration along the remaining noncompact direction. For acceleration along the noncompact direction, the response is stationary and both excitation and de-excitation acquire the same topology-dependent spectral factor, while their ratio continues to satisfy detailed balance at the standard Unruh temperature.

The conventions used throughout are as follows. We employ natural units
$\hbar = c = k_{\mathrm{B}} = 1$ and the metric signature
$(-,+,+,+)$. The detector-field coupling constant is denoted
$\lambda$, and $\varepsilon$ stands for the infinitesimal imaginary
part used in the distributional definition of the Wightman function.

\section{Scalar Field Vacuum on the Two-Torus}
\label{sec:QFT}

We work with four-dimensional Minkowski spacetime, using the coordinates
$(t,x,y,z)$, with the line element
\begin{equation}
ds^2 = -dt^2 + dx^2 + dy^2 + dz^2.
\label{eq:metric}
\end{equation}
The spatial identifications we impose are
\begin{equation}
x \sim x + L_1, \qquad z \sim z + L_2,
\label{eq:identifications}
\end{equation}
where $L_1$ and $L_2$ are the compactification lengths (circumferences of the two circles). The coordinate $y$ remains noncompact and ranges over
$\mathbb{R}$. The resulting spatial topology is
$S^1_{L_1}\times\mathbb{R}\times S^1_{L_2}$. While the metric remains flat everywhere, the periodicity breaks the full
Poincar\'{e} group of the covering space $\mathbb{R}^{3,1}$ as
translations and rotations within the identified directions are
restricted to discrete shifts by multiples of $L_1$ or $L_2$.
These identifications break the full equivalence of inertial frames in special relativity and select a preferred family of inertial frames, namely those in which the compact directions are purely spatial~\cite{PhysRevD.8.1662, Peters:1986at, Uzan:2000wp, Barrow:2001rj}.

We quantize a real, massless scalar field $\phi$
satisfying the Klein-Gordon equation $$\Box\phi = 0$$.
We assume periodic boundary conditions along both compact directions.

The central object we need is the positive-frequency Wightman function
$\WT(x,x') := \braket{0_T|\phi(x)\phi(x')|0_T}$. It is convenient to represent it using the method of images.  On the
universal covering space $\mathbb{R}^{3,1}$, the massless Wightman
function is the well-known
distribution~\cite{Birrell:1982ix, Wald:1995hf, Foo:2022dnz, Bhattacharya:2024jwm}
\begin{equation}
\WM(\Delta x^\mu)
= -\frac{1}{4\pi^2}
\frac{1}{
(\Delta t - i\varepsilon)^2
- (\Delta x)^2
- (\Delta y)^2
- (\Delta z)^2
}.
\label{eq:wightman_mink}
\end{equation}
To apply the periodicity in $x$ with period $L_1$ and in $z$ with period
$L_2$, we have to sum over all images of the
identification~\cite{Davies:1989me}:
\begin{equation}
\WT(x,x')
= \sum_{m=-\infty}^{\infty}
  \sum_{n=-\infty}^{\infty}
  \WM\bigl(
  \Delta t;\;
  \Delta x - mL_1,\;
  \Delta y,\;
  \Delta z - nL_2
  \bigr). 
\label{eq:wightman_torus}
\end{equation}

Subsequently, equation \eqref{eq:wightman_mink} becomes

\begin{equation}
    \WT(x,x')
= \sum_{m=-\infty}^{\infty}
  \sum_{n=-\infty}^{\infty} -\frac{1}{4\pi^2}
\frac{1}{
(\Delta t - i\varepsilon)^2
- (\Delta x-mL_1)^2
- (\Delta y)^2
- (\Delta z-nL_2)^2
}.
\label{eq:explicit_wightman}
\end{equation}

 The set of winding vectors $(mL_1, nL_2)$ forms a
two-dimensional lattice, with norm:
\begin{equation}
\ellmn := \sqrt{(mL_1)^2 + (nL_2)^2},
\label{eq:lattice_norm}
\end{equation}
This norm effectively measures the "topological distance" associated with the
corresponding winding. This lattice norm depends on both
circumferences and will appear throughout our subsequent calculations.

Equation~\eqref{eq:explicit_wightman} provides the starting point for all
the transition-rate computations that follow. Let us record its key properties before proceeding. First, the image sum defines a two-point function that is periodic in $\Delta x$ with period
$L_1$ and in $\Delta z$ with period $L_2$, as required. Second,
when either circumference is sent to infinity the corresponding family
of image terms decouples, reducing~\eqref{eq:explicit_wightman} to the
single-compactification Wightman function studied
in~\cite{Chiou:2016exd}. Third, when both $L_1$ and $L_2$ are taken to
infinity, all images with $(m,n) \neq (0,0)$ vanish, and one recovers
the standard Minkowski two-point
function~\eqref{eq:wightman_mink}. Finally, the local coincidence singularity is governed by the $(0,0)$ image. Non-zero images
can generate additional singularities when the two points are connected by winding null geodesics.

\section{Detector Model and Transition Rate Formalism}
\label{sec:detector}

The Unruh--DeWitt detector~\cite{Unruh:1976db, DeWitt:1980hx, Sanchez:1981xx} is a
conceptually minimal model of a localized quantum system that
interacts with a background field. It consists of a point-like object
carrying two internal energy eigenstates, a ground state $\ket{g}$
and an excited state $\ket{e}$ separated by an energy gap
$\Omega > 0$. The detector is taken to travel along a prescribed worldline
$x^\mu(\tau)$ parameterized by proper time $\tau$. The interaction
between the detector and a real scalar field $\phi$ is governed by the
Hamiltonian
\begin{equation}
H_{\mathrm{int}}(\tau)
= \lambda\,\chi(\tau)\,m(\tau)\,
  \phi\bigl(x^\mu(\tau)\bigr),
\label{eq:interaction}
\end{equation}
where $\lambda$ is a small coupling constant, $m(\tau)$ is the
detector's monopole-moment operator, and $\chi(\tau)$ is a smooth switching function with compact
support that controls when the interaction is turned on and off. The switching function $\chi$ provides the physical
regularization needed to define a finite transition probability in non-equilibrium cases. The role of the switching function and the sharp-switching limit have been analyzed
in detail in \cite{Schlicht:2003iy, Satz:2006kb, Louko:2007mu}.

Using first order perturbation theory in $\lambda$, the amplitude for the detector to transition from $\ket{g}$ to $\ket{e}$, while the field
simultaneously transitions from the vacuum $\ket{0}$ to an excited state $\ket{\psi}$, is given by

\begin{equation}
    i \lambda \bra{\psi,e} \int_{-\infty}^{\infty} \chi(\tau)\,m(\tau)\,
  \phi\bigl(x^\mu(\tau)\bigr) d\tau \ket{0,g},
\end{equation}

Using the interaction-picture evolution $ m(\tau)=e^{iH_0\tau} m(0)e^{-iH_0\tau}$, this amplitude factorizes as

\begin{equation}
    i \lambda \braket{e|m(0)|g}  \int_{-\infty}^{\infty} e^{i \Omega \tau} \chi(\tau)\,\bra{\psi}
  \phi\bigl(x^\mu(\tau)\bigr)\ket{0} d\tau 
  \label{eq:amplitude}
\end{equation}

Using \eqref{eq:amplitude}, the transition probability from $|g\rangle$ to $|e\rangle$ factorizes into a term depending only on the detector’s internal matrix elements $|\braket{e|m(0)|g}|^2$ and the
 response function $F(\Delta E)$ that encodes everything about
the trajectory and the field
state:
\begin{equation}
\mathcal{P}(\Delta E)
= \lambda^2\,
  |\braket{e|m(0)|g}|^2
  \;F(\Delta E),
\label{eq:transition_prob}
\end{equation}
where $\Delta E = +\Omega$ for excitation and
$\Delta E = -\Omega$ for de-excitation, and
\begin{equation}
F(\Delta E)
= \int_{-\infty}^{\infty} d\tau
  \int_{-\infty}^{\infty} d\tau'\;
  e^{-i\Delta E(\tau - \tau')}
  \chi(\tau)\chi(\tau')\;
  \WT\bigl(x(\tau), x(\tau')\bigr).
\label{eq:response_function}
\end{equation}
Our interest lies in the behavior of $F$ and its time derivative for
the three trajectory classes described in Sec.~\ref{sec:intro}.

Two qualitatively different situations arise. If
$\WT\bigl(x(\tau), x(\tau')\bigr)$ depends only on the proper-time
difference $s := \tau - \tau'$ that is, if the pullback of $W_T$ to the worldline depends only on $s$, then the detector reaches equilibrium with the field. In this stationary case one may set $\chi=1$, or equivalently define the rate by dividing the total transition probability by the total proper-time interval, and define the \textit{equilibrium transition
rate}
\begin{equation}
\Feq(\Delta E)
:= \int_{-\infty}^{\infty} ds\;
   e^{-i\Delta E\,s}\;\WT(s),
\label{eq:eq_rate}
\end{equation}
which is simply the Fourier transform of the Wightman
function. Whether a given trajectory satisfies this stationarity
condition depends on the relation between the worldline and the
global identifications: trajectories that are stationary on the
covering space may fail to be so with the
periodicity~\cite{Chiou:2016exd}.

When equilibrium does not hold, the switching function must be
restored, and the physically meaningful quantity is the
\textit{instantaneous transition rate}, obtained by differentiating
$F(\Delta E)$ with respect to the observation time after taking the
sharp-switching limit $\delta/\Delta \to 0$ (where $\delta$ is the
smoothing scale of $\chi$ and $\Delta = \tau - \tau_0$ is the
switch-on duration). Following the regularization-free reformulation
of the response function developed
in~\cite{Satz:2006kb, Louko:2006zv, Louko:2007mu}, the instantaneous rate takes the
form
\begin{equation}
\Finst(\Delta E)
= -\frac{\Delta E}{4\pi}
+ \frac{1}{2\pi^2}
  \int_0^{\Delta} ds
  \left[
  \cos(\Delta E\,s)\;\mathcal{S}(\tau,s)
  + \frac{1}{s^2}
  \right]
+ \frac{1}{2\pi^2 \Delta}
+ O\!\left(\frac{\delta}{\Delta^2}\right),
\label{eq:inst_rate}
\end{equation}
with $$\mathcal{S}(\tau,s) =\; 4\pi^{2}\, W_{T}\!\bigl(x(\tau),\, x(\tau - s)\bigr)\Big|_{\varepsilon = 0}
\;=\; -\sum_{m,n \in \mathbb{Z}}\, \frac{1}{\Delta \sigma^{2}_{mn}(\tau, s)}\bigg|_{\varepsilon = 0},
$$ encodes the full double image sum
evaluated on the trajectory, where $\Delta \sigma^{2}_{mn}(\tau, s)$ denotes the pulled-back squared
geodesic interval between $x(\tau)$ and $x(\tau - s)$, shifted by the
winding vector $(mL_{1}, nL_{2})$, and the $1/s^2$ counterterm cancels the
short-distance divergence inherited from the $(0,0)$ image. 

An important structural identity follows immediately
from~\eqref{eq:inst_rate}: the difference between the de-excitation
and excitation instantaneous rates satisfies
\begin{equation}
\Finst(-|\Delta E|) - \Finst(|\Delta E|)
= \frac{|\Delta E|}{2\pi},
\label{eq:exc_deexc_relation}
\end{equation}
provided $\mathcal{S}(\tau,s)$ has no singularities at nonzero $s$
within the integration range $(0,\Delta)$. This identity was first noted in the context of switched detectors
in \cite{Satz:2006kb} and holds for arbitrary topologies. This relation is topology-independent and follows from the local short-distance structure of the Wightman function, which is unaffected by the global identifications~\cite{Satz:2006kb, Louko:2007mu}. It will serve as a consistency
check in the specific calculations that follow.

\section{Inertial Detector on the Two-Torus}
\label{sec:inertial}

We turn now to the first of our three kinematic settings: a detector
drifting through the toroidal spacetime at constant velocity. Let the
detector's three-velocity be
$\bv = v_z\hat{z}$ (for simplicity we set the
velocity component along $x$ and $y$ to be zero; it can be reinstated
straightforwardly and modifies the results in the expected way). The
worldline is then
\begin{equation}
t = \gamma\tau, \quad
x = \text{constant}, \quad
y = \text{constant}, \quad
z = \gamma v_z \tau,
\label{eq:inertial_worldline}
\end{equation}
with $\gamma = (1 - v_z^2)^{-1/2}$ and four-velocity
$u^\mu = (\gamma, 0, 0, \gamma v_z)$.

Substituting this trajectory into the image-sum Wightman
function~\eqref{eq:wightman_torus} with \eqref{eq:explicit_wightman} now reads as
\begin{equation}
\WT(\tau, \tau^{\prime}) \equiv \WT(s)
= -\frac{1}{4\pi^2}
\sum_{m,n \in \mathbb{Z}}
\frac{1}{
  (u^0 s - i\varepsilon)^2
  - (u^y s)^2
  - (u^z s - nL_2)^2
  - (mL_1)^2
}.
\label{eq:inertial_wightman}
\end{equation}

So the denominator becomes

$$
D_{m n}(s)=s^2+2\left(\gamma v_z n L_2-i \varepsilon \gamma\right) s-\left[\left(m L_1\right)^2+\left(n L_2\right)^2\right] .
$$

This is a quadratic polynomial in (s). Using the quadratic formula, the poles are located at

\begin{equation}
s_{mn}^{(\pm)} = -\gamma v_z nL_2
\pm\mathcal R_{mn} + i\varepsilon\gamma
\label{eq:simple_poles}
\end{equation}

where $
\mathcal R_{mn} = \sqrt{\gamma^2 n^2L_2^2 + m^2L_1^2}
$ is the effective relativistic distance.
For $(m,n)=(0,0)$, we obtain a second-order pole at
$$s=i\varepsilon$$

All other $(m,n)\neq(0,0)$ give two simple poles. The decisive feature is that all poles, both the second-order pole at
$s = i\varepsilon$ and the simple poles $s_{mn}^\pm$, lie in the
upper half of the complex $s$-plane. This follows from the
$i\varepsilon$ prescription and has an immediate consequence for the equilibrium transition rate. For excitation ($\Delta E > 0$), the exponential
$e^{-i\Delta E s}$ decays in the lower half-plane, and the contour is
closed there. Since no poles are enclosed, the contour integral \eqref{eq:eq_rate} vanishes:
\begin{equation}
\Feq(\Delta E) = 0,
\qquad \Delta E > 0.
\label{eq:excitation_zero}
\end{equation}

The vacuum on the two-torus, like the vacuum on the simply-connected
Minkowski spacetime, does not spontaneously excite an inertial
detector at equilibrium. This result is robust against the topology: it
holds for any values of $L_1$ and $L_2$ and for any inertial velocity,
and it reflects the positivity of the field energy as measured in the
preferred frame.

For de-excitation ($\Delta E < 0$), the contour is closed in the
upper half-plane, capturing all the poles. The second-order pole at
$s = i\varepsilon$ produces the standard Minkowski de-excitation rate

\begin{equation}
    \Feq^{(0)}(\Delta E) = 2\pi i \times \text{Res}_{00}
=-\frac{\Delta E}{2\pi}\,\Theta(-\Delta E).
\label{eq:mink_deexcitation}
\end{equation}

The terms with $(m,n)\neq(0,0)$ produce the topological correction to the overall equilibrium transition rate. After evaluating the residues at the simple poles \eqref{eq:simple_poles}, the full 
de-excitation equilibrium rate takes the form

\begin{equation}
\begin{aligned}
\dot{F}(\Delta E)&=2 \pi i\left[\operatorname{Res}_{00}+\sum_{(m, n) \neq(0,0)}\left(\operatorname{Res}_{m n}^{(+)}+\operatorname{Res}_{m n}^{(-)}\right)\right]\\
&=-\frac{\Delta E}{2 \pi}-\frac{i}{4 \pi} \sum_{(m, n) \neq(0,0)} \frac{1}{\mathcal{R}_{m n}}\left[e^{-i \Delta E\left(-\gamma v_z n L_2+\mathcal{R}_{m n}\right)}-e^{-i \Delta E\left(-\gamma v_z n L_2-\mathcal{R}_{m n}\right)}\right]
\label{eq:velocity_derate}
\end{aligned}
\end{equation}

The expression~\eqref{eq:velocity_derate} is not manifestly real as written, but it can be brought into a real form by combining the two complex exponentials. Factoring out the common phase $e^{i\Delta E\,\gamma v_z nL_2}$ and using $e^{-ix}-e^{ix}=-2i\sin x$, the difference in the square brackets becomes $-2i\,\sin(\Delta E\,\mathcal{R}_{mn})\,e^{i\Delta E\,\gamma v_z nL_2}$. The prefactor $-i/(4\pi)$ then combines with $-2i$ to give a factor of $-1/(2\pi)$, and taking the real part replaces the residual exponential by its cosine. The result is
\begin{equation}
\dot{F}(\Delta E)
=-\frac{\Delta E}{2 \pi}
-\frac{1}{2\pi} \sum_{(m, n) \neq(0,0)}
\frac{\sin(\Delta E\,\mathcal{R}_{mn})}{\mathcal{R}_{mn}}
\cos(\Delta E\,\gamma v_z n L_2),
\qquad \Delta E < 0.
\label{eq:velocity_derate_real}
\end{equation}

In the decompactification limit $L_1, L_2 \to \infty$, the winding
amplitude $\Rmn$ diverges for every $(m,n) \neq (0,0)$, the
oscillatory exponentials dephase, and the correction term vanishes,
recovering the standard result. In the
opposite regime of small $L_1$ or $L_2$, the correction becomes
significant and can even dominate the Minkowski term.

The dependence of the correction on $\Rmn$ carries rich physical
information. Because $\Rmn$ involves both $mL_1$ and $nL_2$ in a
quadratic combination, the transition rate is sensitive not only to the
individual circumferences but also to their ratio. Two tori with the same area $L_1L_2$ but different aspect ratios $L_1/L_2$
produce distinguishable detector responses. Furthermore, the rate
depends on the velocity component $v_z$, so an observer moving through the torus will in general measure a different correction from one moving along a single compact direction. In effect, the Unruh-DeWitt detector functions as a
spectrometer for the torus geometry, simultaneously probing both
length scales and the observer's motion relative to the preferred
frame established by the identifications.

The lattice sum in \eqref{eq:velocity_derate_real} is truncated with an isotropic cutoff on the lattice norm, $\ell_{mn} < \Lambda$; details are collected in Appendix~\ref{app:numerics}.

\begin{figure}[h!]
    \centering
    \includegraphics[width=16cm, height=12cm]{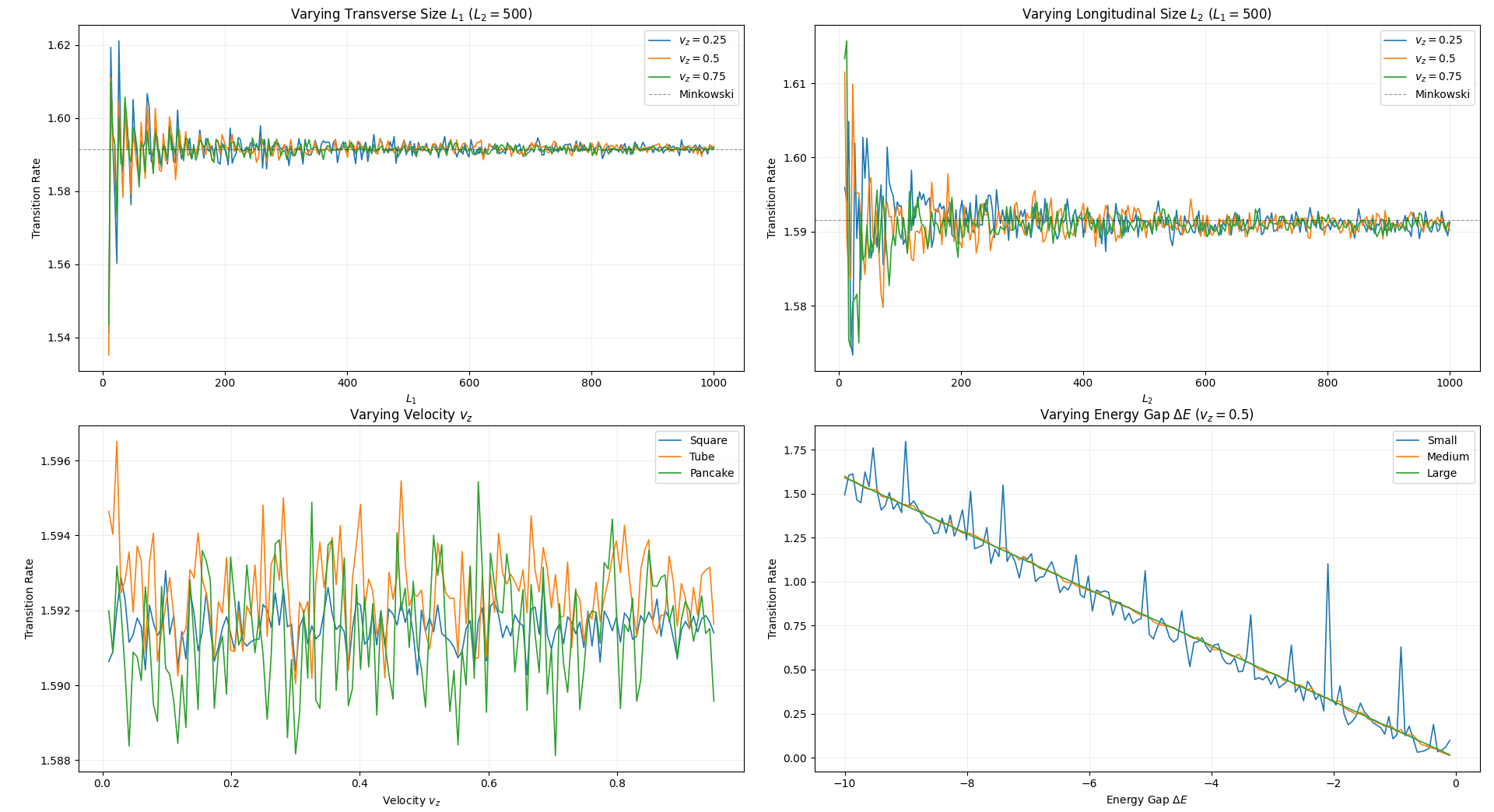}
    \caption{Equilibrium de-excitation rate $\dot{F}_{\rm eq}(\Delta E)$ for an inertial detector on $\mathbb{R} \times T^2$, computed from \eqref{eq:velocity_derate_real} with $\Delta E = -10$. Top left: Rate as a function of $L_1$ at fixed $L_2 = 500$ for three values of the detector velocity $v_z$. Top right: Rate as a function of $L_2$ at fixed $L_1 = 500$ for the same velocities. In both panels, the dashed line marks the Minkowski value $-\Delta E/(2\pi)$. The oscillatory correction decays as the varying circumference increases, and the rate approaches the Minkowski value from above, confirming the decompactification limit. Bottom left: Rate as a function of $v_z$ for three torus aspect ratios (square, tube, and pancake). Bottom right: Rate as a function of $|\Delta E|$ at $v_z = 0.5$ for three torus sizes. The correction is most prominent for the smallest torus and diminishes as $L_1, L_2$ grow.}
    \label{fig:velo_fig}
\end{figure}
The behaviour of the correction term in \eqref{eq:velocity_derate_real} is illustrated in Figure \ref{fig:velo_fig}, computed with $\Lambda = 10\,000$. In the top panels, one of the two circumferences is varied from $10$ to $1000$ while the other is held fixed at $500$. The de-excitation rate oscillates around the Minkowski baseline $-\Delta E/(2\pi) \approx 1.592$ (dashed line) and gradually settles toward it as the varying length increases, which is the expected decompactification behaviour: when either circumference grows, the corresponding family of image terms moves beyond the dominant shells and their contributions become negligible. A notable feature is that the oscillation amplitude and the approach to the asymptotic value depend on the detector velocity $v_z$. At higher velocities, the relativistic distance $R_{mn}$ grows due to the factor of $\gamma$, which pushes the image contributions to larger effective separations and accelerates the convergence to the Minkowski result. This is visible in both panels: the curves for $v_z = 0.75$ settle more quickly than those for $v_z = 0.25$.

The bottom-left panel compares three qualitatively different torus geometries at the same energy gap: a square torus ($L_1 = L_2 = 1000$), an elongated tube configuration ($L_1 = 200, L_2 = 1000$), and a flattened pancake configuration ($L_1 = 1000, L_2 = 200$). The rate depends on the aspect ratio $L_1/L_2$ and not merely on the total area $L_1 L_2$, confirming that the two-dimensional lattice structure carries richer geometric information than a single compactification. The velocity dependence is also different for each geometry, since $v_z$ enters through the winding numbers along the $z$-direction only. The dependence of the correction on the detector velocity vz through the compact
directions is the two-dimensional generalization of the effect first identified in
\cite{Chiou:2016exd} for a single compactification.

The bottom-right panel shows how the de-excitation rate varies with the energy gap $|\Delta E|$ for a fixed velocity $v_z = 0.5$ and three torus sizes: small $L_1 = L_2 = 10$, medium $L_1 = L_2 = 100$, and large $L_1 = L_2 = 1000$. For the smallest torus, the correction is large and produces visible oscillations at all energy scales, while for the largest torus, the rate is nearly indistinguishable from the Minkowski result. This confirms that a measurement of $\dot{F}_{\rm eq}(\Delta E)$ over a range of energy gaps can, in principle, be used to extract both compactification lengths and the detector's velocity relative to the preferred frame. 

\begin{figure}[h!]
    \centering
    \includegraphics[width=16cm]{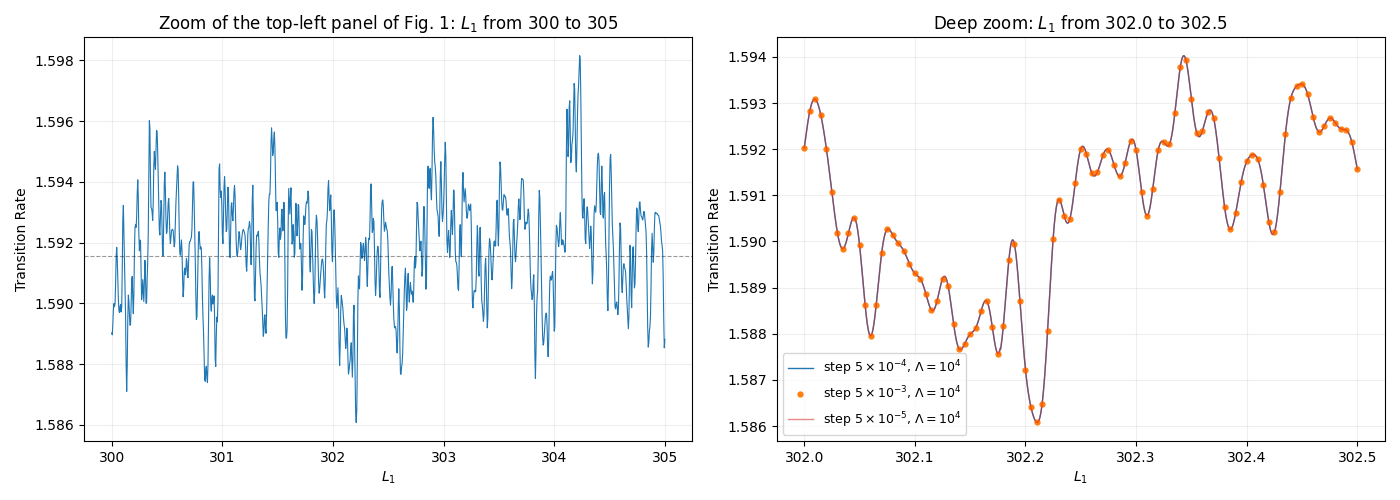}
    \caption{Zoom of the top-left panel of Figure~\ref{fig:velo_fig}, computed from \eqref{eq:velocity_derate_real} with the same parameters and the same radial cutoff $\Lambda = 10^4$; only the plot window and the sampling step differ. \textit{Left:} $L_1$ from $300$ to $305$ at step $5\times10^{-3}$. \textit{Right:} $L_1$ from $302.0$ to $302.5$ at steps $5\times10^{-3}$, $5\times10^{-4}$ and $5\times10^{-5}$. The three curves coincide, and the largest difference between adjacent samples falls from $2.0\times10^{-3}$ to $2.1\times10^{-4}$ to $4.3\times10^{-5}$, in proportion to the step. The zooms establish that the displayed finite-cutoff curves
are continuous and are not sampling artefacts. We have additionally checked their stability under
changes of the lattice regulator, as described in Appendix \ref{app:numerics}.}
    \label{fig:zoomL1}
\end{figure}

\begin{figure}[h!]
    \centering
    \includegraphics[width=16cm]{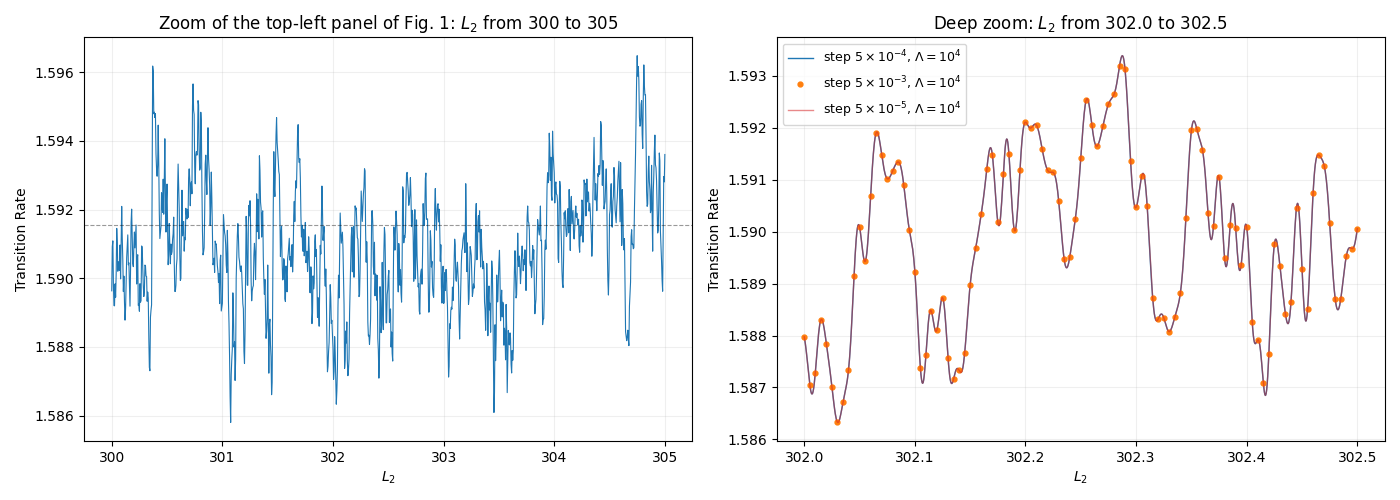}
    \caption{As Figure~\ref{fig:zoomL1}, for the top-right panel of Figure~\ref{fig:velo_fig}, with $L_2$ varied at fixed $L_1 = 500$. \textit{Left:} $L_2$ from $300$ to $305$ at step $5\times10^{-3}$. \textit{Right:} $L_2$ from $302.0$ to $302.5$ at steps $5\times10^{-3}$, $5\times10^{-4}$ and $5\times10^{-5}$. The largest difference between adjacent samples falls from $2.2\times10^{-3}$ to $2.6\times10^{-4}$ to $2.6\times10^{-5}$.}
    \label{fig:zoomL2}
\end{figure}

\section{Uniformly Accelerated Detector Along a Compact Direction}
\label{sec:accel_compact}

We now consider a detector undergoing constant proper acceleration
$a = 1/\alpha$ directed along $z$, one of the two compact directions.
The trajectory is the standard Rindler hyperbola:
\begin{equation}
t(\tau) = \alpha\sinh\frac{\tau}{\alpha}, \quad
x = 0, \quad
y = 0, \quad
z(\tau) = \alpha\cosh\frac{\tau}{\alpha}.
\label{eq:compact_accel_worldline}
\end{equation}
This worldline is an orbit of the boost Killing vector in the Rindler wedge, and in the covering
space $\mathbb{R}^{3,1}$ the Wightman function would
depend only on $s = \tau - \tau'$. On the identified spacetime,
however, the situation changes fundamentally.

Inserting~\eqref{eq:compact_accel_worldline} into the image-sum
formula~\eqref{eq:explicit_wightman} and evaluating each Minkowski image,
we find

\begin{equation}
\begin{split}
&\WT(\tau,\tau') = \\& -\frac{1}{4\pi^2} \sum_{m,n} \frac{1}{\Biggl[ \left( \alpha\sinh\frac{\tau}{\alpha} - \alpha\sinh\frac{\tau'}{\alpha} - i\varepsilon \right)^2 
 - \left( \alpha\cosh\frac{\tau}{\alpha} - \alpha\cosh\frac{\tau'}{\alpha} - nL_2 \right)^2 - (mL_1)^2 \Biggr] }.
\end{split}
\label{eq:compact_wightman}
\end{equation}
To expose the dependence on individual proper times, we introduce the
half-sum $\Sigma := (\tau + \tau')/2$ and difference
$s := \tau - \tau'$ and apply standard hyperbolic addition formulas.
The denominator of the $(m,n)$ summand becomes the invariant
separation
\begin{equation}
\Delta\sigma^2_{mn}(\Sigma, s)
= 4\alpha^2\sinh^2\frac{s-i\varepsilon}{2\alpha}
-4\alpha nL_2
\sinh\frac{\Sigma}{\alpha}
\sinh\frac{s}{2\alpha}- n^2L_2^2 - m^2L_1^2.
\label{eq:compact_invariant}
\end{equation}

When both compact lengths are much larger than the acceleration scale: $L_1, L_2 \gg \alpha$, all images with $(m,n)\neq(0,0)$ decouple. Only the $(m,n)=(0,0)$ term survives. Thus, \eqref{eq:compact_invariant} reduces to

$$\Delta\sigma^2_{mn}(\Sigma, s)
= 4\alpha^2\sinh^2\frac{s-i\varepsilon}{2\alpha}$$

Subsequently, the Wightman function \eqref{eq:compact_wightman} reduces to,
\begin{equation}
 \WT(\tau,\tau') = -\frac{1}{16\pi^2\alpha^2} \frac{1}{\sinh^2\left(\frac{\tau-\tau'-i\varepsilon}{2\alpha}\right)}
\end{equation}
This depends only on $s=\tau-\tau'$, so equilibrium holds.
Using the standard contour argument:
$$\dot F(\Delta E) =\frac{\Delta E}{2\pi}
\frac{1}{e^{2\pi\alpha\Delta E}-1}.$$
This is the standard Unruh result.

The $\Sigma$-dependence shows that the Wightman function is
non-stationary: the two-point function depends
on both $\tau$ and $\tau'$ individually, not just on their difference. Because the stationarity condition~\eqref{eq:eq_rate} is violated, the
equilibrium transition rate is not defined for this trajectory. One
must instead work with the instantaneous
rate~\eqref{eq:inst_rate}. The image-sum factor entering that formula
is
\begin{equation}
\mathcal{S}(\tau,s)
= -\sum_{m,n}
\frac{1}{\Delta\sigma^2_{mn}(\tau,s)}
\bigg|_{\varepsilon=0} = 4 \pi^2 \WT(\tau,\tau-s)\bigg|_{\varepsilon=0},
\label{eq:compact_S}
\end{equation}
yielding
\begin{equation}
\label{eq:acc_instantrate}
\begin{split}
    \Finst(\Delta E) = & -\frac{\Delta E}{4\pi} + \frac{1}{2\pi^2}\int_0^\Delta ds \Biggl[ \cos(\Delta E s) \\
    & \times -\sum_{m,n}\frac{1}{4\alpha^2\sinh^2\frac{s}{2\alpha}- 4\alpha nL_2\sinh\frac{2\tau-s}{2\alpha}\sinh\frac{s}{2\alpha}-n^2L_2^2-m^2L_1^2} \\
    & + \frac{1}{s^2} \Biggr] +\frac{1}{2\pi^2\Delta} + O\left(\frac{\delta}{\Delta^2}\right).
\end{split}
\end{equation}

with $\Sigma$ set to $\tau - s/2$ (since
$\Sigma = \tau - s/2$ when $\tau' = \tau - s$). 

Equation~\eqref{eq:acc_instantrate} must be evaluated numerically: the integrand has no closed-form antiderivative, and moreover it develops singularities within the integration domain at the critical values of~$s$ discussed below. The lattice modes $(m,n)$ entering the sum are selected by a radial cutoff $\ell_{mn} < \Lambda$ (with $\Lambda = 500$ in units of $\Delta E$) and sorted by increasing $\ell_{mn}$, following the same shell-ordered protocol described in Section~\ref{sec:inertial}. At $s = 0$, the $(0,0)$ piece of~$\mathcal{S}(\tau,s)$ diverges as $-1/s^2$, which is exactly cancelled by the $+1/s^2$ counterterm. We impose this analytic limit for $s < 10^{-4}$ to avoid numerical cancellation. The integration from $0$ to~$\Delta$ is then carried out using adaptive Gaussian quadrature with absolute and relative tolerances of~$10^{-5}$. The decompactification limit is recovered by setting the relevant circumference to $10\,000/\Delta E$, which is sufficiently large that the corresponding winding modes fall outside the radial cutoff. Further details of the numerical procedure, including the critical-time computation described below, are collected in Appendix~\ref{app:numerics}.

A distinctive feature of the two-torus background is the structure of
the singularities of $\mathcal{S}(\tau,s)$. A singularity arises
whenever $\Delta\sigma^2_{mn} = 0$ for some
$(m,n) \neq (0,0)$ at a nonzero value of $s$. The origin of these singularities is the light-cone singularity of the free two-point function. By construction $\Delta\sigma^2_{mn}$ is the squared Minkowski interval, on the covering space, between $x(\tau)$ and the image of $x(\tau-s)$ displaced by the winding vector, and the corresponding summand of \eqref{eq:explicit_wightman} is singular exactly where that interval vanishes. On the quotient the image is not a separate event but the same event, reached along a geodesic that winds $m$ times around the $x$-circle and
$n$ times around the $z$-circle. The condition $\Delta\sigma^2_{mn} = 0$ therefore states that a null geodesic of the identified spacetime joins two distinct points of the detector's own worldline. In Minkowski space a timelike worldline meets its own light cone only at $s = 0$, and the coincidence pole is the only one; the identifications place further null contacts at finite $s$, and each contributes a pole. 

\begin{figure}[h!]
    \centering
    \includegraphics[width=15cm, height=6cm]{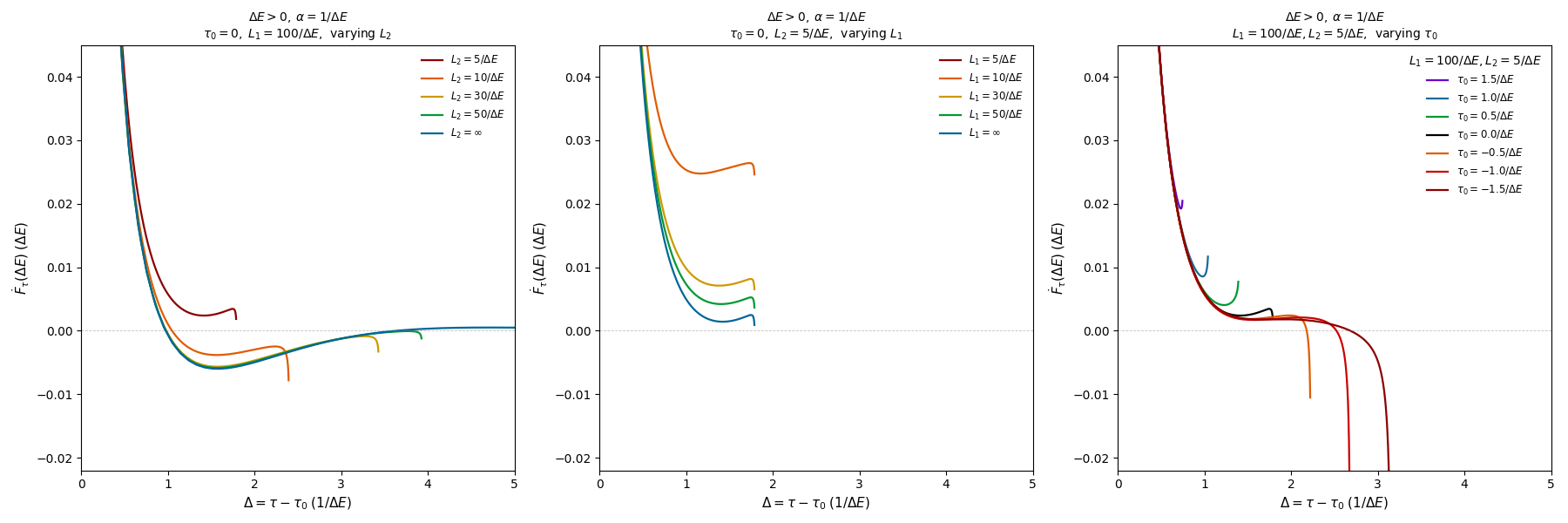}
    \caption{Instantaneous excitation rate $\dot{F}_\tau(\Delta E)$ as a function of the switch-on duration $\Delta = \tau - \tau_0$ for the trajectory \eqref{eq:compact_accel_worldline} with $\alpha = 1/\Delta E$ and $\Delta E > 0$. \textit{Left:} $L_1 = 100/\Delta E$ fixed, $\tau_0 = 0$, varying $L_2$. \textit{Middle:} $L_2 = 5/\Delta E$ fixed, $\tau_0 = 0$, varying $L_1$. \textbf{Right:} $L_1 = 100/\Delta E$, $L_2 = 5/\Delta E$ fixed, varying $\tau_0$. The $L = \infty$ curves approach the Unruh equilibrium value asymptotically, while all finite-$L$ curves diverge at the critical duration $s_c^{\min}$ given by \eqref{eq:27}-\eqref{eq:critical}.}
    \label{fig:Figsection5}
\end{figure}

To find these critical durations explicitly we set $\Delta\sigma^2_{mn} = 0$ at $\epsilon = 0$ in \eqref{eq:compact_invariant} and substitute $\Sigma = \tau_0 + \Delta - s/2$ (since $\tau = \tau_0 + \Delta$ and $\tau' = \tau - s$). They matter because \eqref{eq:inst_rate} assumes the integrand is regular on the whole of $(0,\Delta)$, so it applies only for switch-on durations shorter than the smallest such $s$, and those durations are where the curves of Figure~\ref{fig:Figsection5} diverge. Writing $X := \sinh(s_c/2\alpha)$, the vanishing condition becomes the quadratic

\begin{equation}
   4\alpha^2 X^2 + 4\alpha n L_2 \sinh\!\left(\frac{\tau_0 + \Delta - s_c/2}{\alpha}\right) X = n^2 L_2^2 + m^2 L_1^2 \equiv \ell^2_{mn}, 
   \label{eq:27}
\end{equation}

which can be solved formally as

\begin{equation}
    X = \frac{-K + \sqrt{K^2 + \ell^2_{mn}/\alpha^2}}{2}, \qquad K := \frac{nL_2}{\alpha}\sinh\!\left(\frac{\tau_0 + \Delta - s_c/2}{\alpha}\right). 
    \label{eq:critical}
\end{equation}

The critical duration is then recovered from $s_c^{(m,n)} = 2\alpha\,\mathrm{arcsinh}(X)$. Because $K$ itself depends on $s_c$ through the argument of the sinh, equations \eqref{eq:27} - \eqref{eq:critical} must therefore be solved implicitly. Note that since $\tau = \tau_0 + \Delta$ and $\tau' = \tau - s$, we have $\Sigma = \tau_0 + \Delta - s/2$; when the integration variable is the critical duration itself ($s = s_c$, with $\Delta$ being the upper limit of integration), the argument of $K$ can equivalently be written as $(\tau_0 + s_c/2)/\alpha$, as used in Appendix~\ref{app:numerics}.

Two limiting cases are worth noting. First, when $L_1 \to \infty$ with $L_2$ fixed, only the $m = 0$ modes contribute, the lattice norm reduces to $\ell_{0n} = |n|L_2$, and \eqref{eq:27} reproduces the single-compactification critical condition of \cite{Chiou:2016exd}. Second, as $\tau_0$ increases, the first winding-null intersection occurs at a smaller
proper-time separation $s_c$.

With a single compact dimension, these critical times form a
one-dimensional sequence indexed by a single winding number. With two
compact dimensions, they form a two-dimensional set
\begin{equation}
\{s_c^{(m,n)} : (m,n) \in \mathbb{Z}^2 \setminus \{(0,0)\}\},
\label{eq:critical_set}
\end{equation}
ordered by the lattice norm $\ellmn$. The smallest critical time,
\begin{equation}
s_c^{\min}
= \min_{(m,n) \neq (0,0)} s_c^{(m,n)},
\label{eq:critical_min}
\end{equation}
determines the maximum switch-on duration for which the
formula~\eqref{eq:inst_rate} remains valid. Because the
two-dimensional lattice has a higher density of short vectors than any
one-dimensional sub-lattice, this safe observation window is
generically \textit{shorter} than in the single-compactification case
at comparable length scales. Although the instantaneous rate diverges as $\Delta\to s_c^{\min}$, the divergence is integrable, so the total transition probability remains finite.

Several features of Figure~\ref{fig:Figsection5} are worth highlighting. All three panels are computed with the lattice sum truncated at $\ell_{mn} < 500/\Delta E$, and the integration is carried out from $\Delta = 0.02/\Delta E$ up to a value $\Delta_{\max} = s_c^{\min} - 0.005/\Delta E$, just short of the first critical time determined numerically from \eqref{eq:27}--\eqref{eq:critical}. The curves are therefore exact within the plotted range and diverge only because the integrand itself blows up as $\Delta \to s_c^{\min}$.

In the left panel, $L_1 = 100/\Delta E$ is held fixed while $L_2$ is varied. Reducing $L_2$ from $\infty$ (computed numerically by setting $L_2 = 10\,000/\Delta E$) to $5/\Delta E$ pulls the divergence to progressively smaller values of $\Delta$, confirming that the safe observation window shrinks as the compact direction along which the detector accelerates becomes smaller. The $L_2 = \infty$ curve is effectively the single-compactification result with only $L_1$ compact; it does not diverge within the plotted range because $L_1 = 100/\Delta E$ is large enough that the first critical time exceeds $\Delta = 5/\Delta E$. As $L_2$ decreases, the $(0, n)$ winding modes with small $|n|$ produce shorter return times, and the divergence moves to the left. The onset of the divergence is steep but integrable, consistent with the general expectation that the total transition probability remains finite even though the instantaneous rate blows up.

The middle panel shows the complementary situation: $L_2 = 5/\Delta E$ is held fixed while $L_1$ is varied. The position of the divergence is now set primarily by $L_2$ and is only weakly affected by $L_1$, because the acceleration is along the $z$-direction and the null signals that return fastest are those that wind around the $z$-circle (the $(0, \pm 1)$ modes) rather than the $x$-circle. When $L_1$ is also made small, the additional $(m, 0)$ and mixed $(m, n)$ modes bring their own critical times into the observation window, and the curve begins to deviate earlier. The $L_1 = \infty$ curve still diverges, because the $z$-compactification alone is sufficient to generate topological echoes.

The right panel reveals the dependence on the switch-on time $\tau_0$ at fixed $L_1 = 100/\Delta E$, $L_2 = 5/\Delta E$. Varying $\tau_0$ is equivalent to changing the detector's instantaneous velocity $u^z(\tau_0) = \sinh(\tau_0/\alpha)$ at the moment it is turned on. As $\tau_0$ increases, the first winding-null intersection occurs at a smaller proper-time separation $s_c$: numerically, $s_c^{\min}$ decreases from approximately $2.8/\Delta E$ at $\tau_0 = -1.5/\Delta E$ to approximately $1.2/\Delta E$ at $\tau_0 = 1.5/\Delta E$, pulling the divergence to smaller $\Delta$. This $\tau_0$-dependence has no counterpart in the decompactified case ($L_1, L_2 \to \infty$), where the instantaneous rate is independent of $\tau_0$ and asymptotes to the equilibrium Unruh value.

\section{Uniformly Accelerated Detector in the Noncompact Direction}
\label{sec:accel_noncompact}

The third setting we consider is a detector accelerating with constant
proper acceleration $a = 1/\alpha$ along $y$, the non-compact
direction. The worldline reads
\begin{equation}
t(\tau) = \alpha\sinh\frac{\tau}{\alpha}, \quad
x = 0, \quad
y(\tau) = \alpha\cosh\frac{\tau}{\alpha}, \quad
z = 0.
\label{eq:noncompact_worldline}
\end{equation}
Inserting this trajectory into~\eqref{eq:explicit_wightman}, the
pulled-back Wightman function takes the form
\begin{equation}
\WT(\tau,\tau')
= -\frac{1}{4\pi^2}
\sum_{m,n}
\frac{1}{
\left(
\alpha\sinh\frac{\tau}{\alpha}
- \alpha\sinh\frac{\tau'}{\alpha}
- i\varepsilon
\right)^2
- \left(
\alpha\cosh\frac{\tau}{\alpha}
- \alpha\cosh\frac{\tau'}{\alpha}
\right)^2
- (mL_1)^2
- (nL_2)^2
}.
\label{eq:noncompact_wightman}
\end{equation}
A critical simplification now occurs. The combination
\begin{equation}
\left(
\alpha\sinh\frac{\tau}{\alpha}
- \alpha\sinh\frac{\tau'}{\alpha}
\right)^2
- \left(
\alpha\cosh\frac{\tau}{\alpha}
- \alpha\cosh\frac{\tau'}{\alpha}
\right)^2
= -4\alpha^2 \sinh^2\!\frac{s}{2\alpha}
\label{eq:hyperbolic_identity}
\end{equation}
depends solely on $s = \tau - \tau'$, and the image shifts
$(mL_1)^2 + (nL_2)^2$ are constant. Therefore the full
expression~\eqref{eq:noncompact_wightman} reduces to a function of $s$
alone:
\begin{equation}
\WT(s)
= -\frac{1}{16\pi^2\alpha^2}
\sum_{m,n}
\frac{1}{
\sinh^2\!\left(
\frac{s}{2\alpha} - \frac{i\varepsilon}{2\alpha}
\right)
- \frac{\ell_{mn}^2}{4\alpha^2}
},
\label{eq:noncompact_wightman_simplified}
\end{equation}
where we have already introduced the lattice norm $\ell_{mn}$, defined in eq.\eqref{eq:lattice_norm}.

The stationarity condition is satisfied: the detector is in equilibrium
with the field, and the transition rate can be obtained
from~\eqref{eq:eq_rate} by contour integration. The reason for this
restored equilibrium is geometrical. The
worldline~\eqref{eq:noncompact_worldline} generates the boost Killing
vector $y\partial_t + t\partial_y$, which commutes with translations
in $x$ and $z$. The identifications~\eqref{eq:identifications} along
$x$ and $z$ therefore do not break the symmetry generated by this
boost, and the trajectory remains stationary on the quotient
spacetime, In contrast to the case treated in
Sec.~\ref{sec:accel_compact}, where the acceleration was aligned with
a compact direction.

To compute the transition rate we analyze the poles
of~\eqref{eq:noncompact_wightman_simplified}. Because $\sinh^2$ is
periodic under $s \to s + 2\pi i\alpha$, the pole set of each summand
is invariant under that shift, so the poles are organized into rows
stacked at vertical spacing $2\pi\alpha$. For the $(0,0)$ image,
$\ell_{00} = 0$, and the singularities are at
$\sinh^2(s/2\alpha) = 0$, i.e., $s = 2\pi ik\alpha + i\varepsilon$ with
$k \in \mathbb{Z}$; these are the double poles of the standard
Rindler Wightman function. For
$(m,n) \neq (0,0)$, the condition
$\sinh^2(s/2\alpha) = \frac{\ell_{mn}^2}{4\alpha^2}$ yields
\begin{equation}
s_{mn}^{\pm,k}
= \pm 2\alpha\,\arcsinh\,\frac{\ell_{mn}}{2\alpha} + i\bigl(2\pi\alpha k + \varepsilon\bigr),
\qquad k \in \mathbb{Z},
\label{eq:noncompact_poles}
\end{equation}
a pair of simple poles in every row. The rows with $k \geq 0$ lie in the
upper half $s$-plane and those with $k \leq -1$ in the lower half, so
each channel encloses one half of the same ladder. Note that the image
poles are not confined to the upper half-plane: only the $k = 0$ row of
\eqref{eq:noncompact_poles} lies just above the real axis, and the rows
below it contribute to the excitation channel. While analysing Eq. \eqref{eq:noncompact_poles}, we found that Eq. (8.6) of Ref. \cite{Chiou:2016exd} is incomplete because it omits the $+2\pi i \alpha k$ term required for the full pole ladder. This omission also removes the image poles below the real axis and hence their contribution to the excitation channel.

The row structure makes the two channels easy to compare. Shifting a
row upward by $2\pi i\alpha k$ leaves $\WT$ unchanged and multiplies
$e^{-i\Delta E s}$ by $e^{2\pi\alpha\Delta E k}$, so the total residue of
row $k$ is that of row $0$ times $q^k$ with $q = e^{2\pi\alpha\Delta E}$.
Writing the row-$0$ contribution as
\begin{equation}
\mathcal{B}(\Delta E)
= \frac{\Delta E}{2\pi}
+ \Corr(\Delta E, \alpha, L_1, L_2),
\label{eq:row_zero}
\end{equation}
where the topological part is
\begin{equation}
\displaystyle
\Corr(\Delta E, \alpha, L_1, L_2)
= \sum_{(m,n) \neq (0,0)}
\frac{
\sin\bigl(
2\alpha\Delta E\,\arcsinh\,\frac{\ell_{mn}}{2\alpha}
\bigr)
}{
2\pi\,
\ell_{mn}\sqrt{1 + \bigl(\frac{\ell_{mn}}{2\alpha}\bigr)^2}
},
\label{eq:topo_correction}
\end{equation}
the geometric sums over the two halves of the ladder differ only in
that the upward sum starts one row earlier. Summing them
(see Appendix~\ref{app:contour}) gives a single expression valid for
both signs of $\Delta E$,
\begin{equation}
\Feq(\Delta E)
= \frac{1}{e^{2\pi\alpha\Delta E} - 1}
\left[
\frac{\Delta E}{2\pi}
+ \Corr(\Delta E, \alpha, L_1, L_2)
\right].
\label{eq:noncompact_rate}
\end{equation}
In particular, for excitation
\begin{equation}
\Feq(\Delta E)
= \frac{\mathcal{B}(\Delta E)}{e^{2\pi\alpha\Delta E} - 1},
\qquad
\Feq(-\Delta E)
= \frac{e^{2\pi\alpha\Delta E}\,\mathcal{B}(\Delta E)}{e^{2\pi\alpha\Delta E} - 1},
\qquad \Delta E > 0,
\label{eq:excitation_unruh}
\end{equation}
so that the topological correction enters both channels through the same
bracket and cancels in the ratio:
\begin{equation}
\frac{\Feq(\Delta E)}{\Feq(-\Delta E)}
= e^{-2\pi\alpha\Delta E}.
\label{eq:detailed_balance}
\end{equation}
The detector therefore satisfies the detailed-balance form of the KMS
condition at the Unruh temperature $T = 1/(2\pi\alpha)$, and the
temperature it reports carries no dependence on $L_1$ or $L_2$. The
reason is visible in the pole structure: the temperature is fixed by the
vertical spacing $2\pi\alpha$ of the ladder, which depends only on the
acceleration, whereas the identifications displace poles horizontally
within a row. Compactification changes how often the detector clicks,
through $\Corr$, but not the temperature it measures.

Several properties
of~\eqref{eq:topo_correction}--\eqref{eq:noncompact_rate}
deserve comment. First, in the decompactification limit
$L_1, L_2 \to \infty$, $\ell_{mn} \to \infty$ for all
$(m,n) \neq (0,0)$ and $\Corr \to 0$; one recovers the pure Unruh
result for both excitation and de-excitation. Second, sending $L_1 \to \infty$ with $L_2$ fixed eliminates all $m \neq 0$ terms and reduces
~\eqref{eq:topo_correction} to a single sum over $n$, the single-compactification case, with the full pole ladder of \eqref{eq:noncompact_poles} retained. Third, the correction depends on the individual lengths $L_1$ and $L_2$, not
simply on their product: tori of equal area but different aspect ratios
yield different de-excitation rates. This aspect-ratio sensitivity is a
genuinely two-dimensional effect with no counterpart in the
single-compactification setting. Fourth, the zero-acceleration limit
$\alpha \to \infty$
reduces~\eqref{eq:noncompact_rate} to the inertial
result~\eqref{eq:excitation_zero}-\eqref{eq:velocity_derate_real}
evaluated at zero velocity: the prefactor tends to $-1$ for
$\Delta E < 0$, while $\arcsinh(\ell_{mn}/2\alpha) \to \ell_{mn}/2\alpha$
turns each summand of \eqref{eq:topo_correction} into
$\sin(\Delta E \ell_{mn})/(2\pi\ell_{mn})$. Fifth, as $\ell_{mn} \to 0$ a
summand of \eqref{eq:topo_correction} tends to $\Delta E/(2\pi)$, matching
the $(0,0)$ term, as it must since the two simple poles merge into the
Rindler double pole in that limit.

\begin{figure}[h!]
    \centering
    \includegraphics[width=16cm, height=12cm]{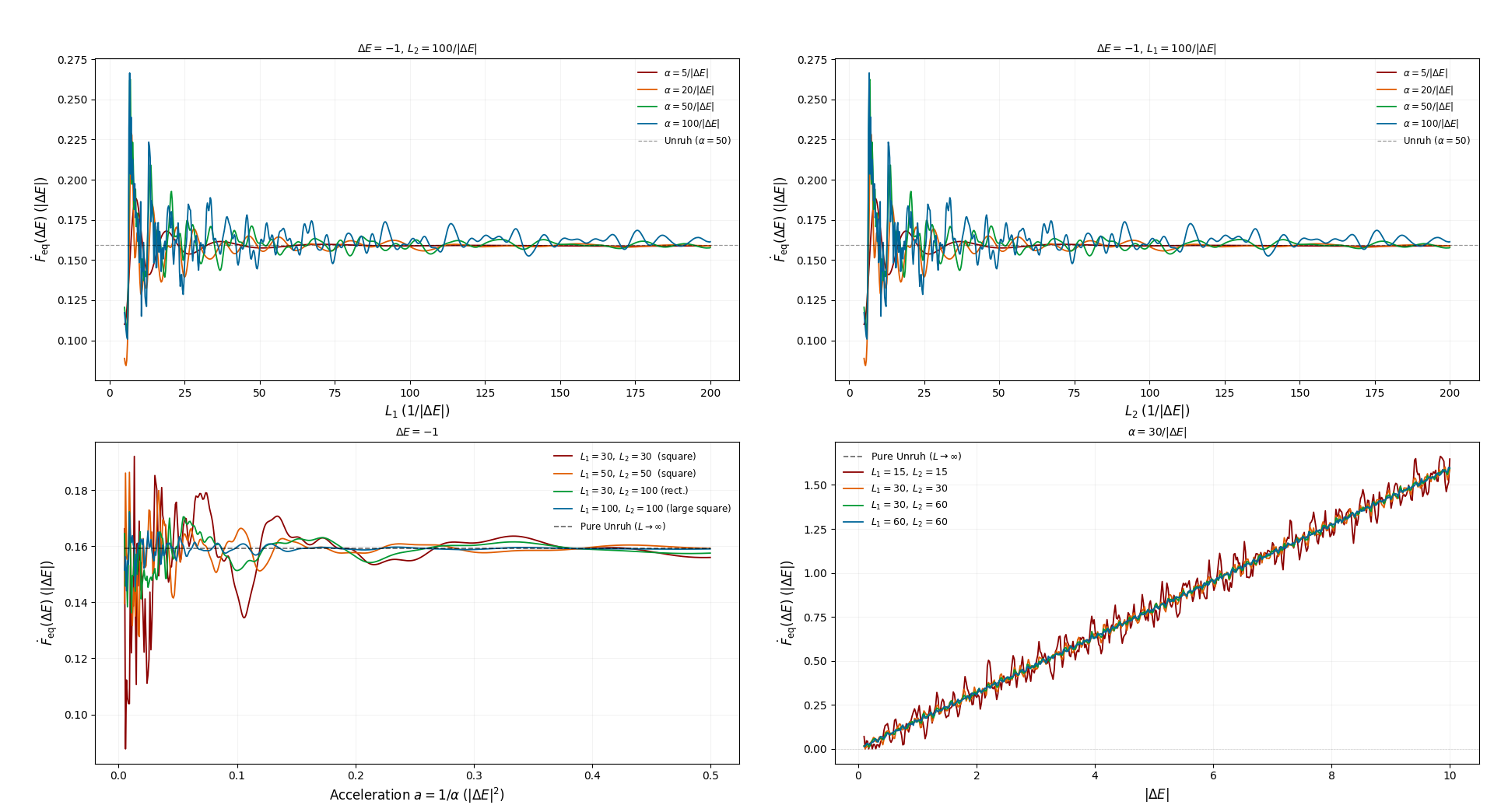}
    \caption{Equilibrium de-excitation rate $\dot{F}_{\rm eq}(\Delta E)$ for a detector accelerating along the noncompact direction, computed from \eqref{eq:noncompact_rate}. All quantities are expressed in units of $|\Delta E|$. \textit{Top left:} Rate as a function of $L_1$ at fixed $L_2 = 100/|\Delta E|$ and $\Delta E = -1$, for four values of $\alpha$. \textit{Top right:} Rate as a function of $L_2$ at fixed $L_1 = 100/|\Delta E|$, same parameters. The dashed line indicates the pure Unruh value for $\alpha = 50/|\Delta E|$. $\textit{Bottom left:}$ Rate as a function of the proper acceleration $a = 1/\alpha$ at fixed $\Delta E = -1$, for four torus geometries including a rectangular case. The dashed line is the pure Unruh result ($L \to \infty$). \textit{Bottom right:} Rate as a function of $|\Delta E|$ with $\alpha |\Delta E| = 30$ held fixed, for four torus configurations. The dashed line is again the pure Unruh spectrum.}
    \label{fig: Figsect6}
\end{figure}

The behaviour of the de-excitation rate \eqref{eq:noncompact_rate} is shown in Figure \ref{fig: Figsect6}, computed with $\Lambda = 1\,000/|\Delta E|$. In the top two panels, one circumference is varied from $5/|\Delta E|$ to $200/|\Delta E|$ while the other is held fixed at $100/|\Delta E|$, and the rate is plotted for four values of the acceleration parameter $\alpha$. At small values of the varying length the topological correction $\mathcal{C}$ is large, and the rate oscillates significantly around the pure Unruh baseline (dashed line, shown for $\alpha = 50/|\Delta E|$). As the length increases, the relevant image contributions decouple and the rate converges to the Unruh value, confirming the decompactification limit. The oscillation envelope decays on a scale set by $\alpha$: for $\alpha = 5/|\Delta E|$ the rate has already settled by $L \sim 20/|\Delta E|$, while for $\alpha = 100/|\Delta E|$ the oscillations persist well beyond $L = 100/|\Delta E|$. This is a direct consequence of the structure of the summand in \eqref{eq:noncompact_rate}, whose amplitude for the leading $(m,n)$ mode scales as $\sim 1/(2\pi L)$ when $L \ll 2\alpha$ and as $\sim \alpha/(\pi L^2)$ when $L \gg 2\alpha$; the crossover between these regimes occurs at $L \sim 2\alpha$. Comparing the two top panels, the rate depends on $L_1$ and $L_2$ in a structurally identical way when the other length is large, which reflects the symmetric role played by the two circumferences when the acceleration lies along the noncompact direction. Unlike Figure~\ref{fig:Figsection5}, no divergence appears in any panel, because the poles of the pulled-back Wightman function are displaced from the real axis and the trajectory remains stationary.

The bottom-left panel shows the rate as a function of the proper acceleration $a = 1/\alpha$ for four torus geometries. At large acceleration (right side of the panel) the Unruh thermal contribution dominates the de-excitation rate and all curves converge to the pure Unruh result (dashed line), which itself grows as $a$ increases. At small acceleration the topological correction becomes comparable to the Unruh term and the individual torus geometries separate. A particularly informative comparison is between the $L_1 = 30, L_2 = 100$ rectangular torus and the $L_1 = 50, L_2 = 50$ square torus: both have comparable area ($3000$ vs $2500$ in units of $1/|\Delta E|^2$), yet they produce clearly distinguishable curves. This confirms that the correction \eqref{eq:noncompact_rate} depends on the individual lengths $L_1$ and $L_2$ and not simply on their product, so that tori of equal area but different aspect ratios are distinguishable.

The bottom-right panel displays the spectral dependence of the de-excitation rate along the line $\alpha |\Delta E| = 30$. Since $|\Delta E|$ is varied across this panel, $\alpha$ itself is not held fixed. The pure Unruh spectrum (dashed line) is nearly linear at large $|\Delta E|$, approaching $|\Delta E|/(2\pi)$. The topological correction superimposes oscillations whose amplitude grows with $|\Delta E|$ for the smallest torus ($L_1 = L_2 = 15/|\Delta E|$, dark red) and is barely visible for the largest ($L_1 = L_2 = 60/|\Delta E|$, blue). The oscillation period in $|\Delta E|$ is governed by the argument of the sine in \eqref{eq:noncompact_rate}, which is approximately $|\Delta E| \cdot \ell_{\rm min}$ for the leading mode; a smaller torus therefore produces slower, larger oscillations. At sufficiently large $|\Delta E|$ the oscillations from multiple lattice modes interfere and the curve acquires an irregular appearance, but the correction remains bounded and the rate never diverges, in contrast to the situation in Section~\ref{sec:accel_compact} where the acceleration is along a compact direction. This regularity is a direct consequence of the equilibrium character of the trajectory.

\section{Discussion and Outlook}
\label{sec:discussion}

The results obtained in this paper demonstrate that a local
Unruh--DeWitt detector is, in principle, capable of distinguishing the
spatial topology $\mathbb{R}\times T^2$ from both $\mathbb{R}^3$ and
$\mathbb{R}^2\times S^1$, and moreover of extracting quantitative
information about the geometry of the torus, its two circumferences
and their ratio.

For the inertial detector, the key signature is a non-vanishing correction to the equilibrium de-excitation rate that depends on both $L_1$ and $L_2$ as well as the detector's velocity components along the compact directions. The excitation rate remains identically zero. This shows that the topology does not destabilize the vacuum. Instead, topology redistributes the spectral weight of vacuum fluctuations, and that redistribution appears only in the de-excitation channel. The dual dependence on $L_1$ and $L_2$ means that an observer who measures the de-excitation rate over a range of energy gaps can, in principle, disentangle two independent length scales. In a more general inertial setup with motion along both compact directions, the rate should also depend on two velocity components. This is a significantly richer diagnostic capability than is available with a single compactification. It is also worth noting that the correction in \eqref{eq:velocity_derate_real} depends on the relativistic distance $\Rmn$, which mixes the two circumferences through the Lorentz factor $\gamma$.

For the detector accelerating along a compact direction, the principal new feature relative to the single-compactification case is the organization of the critical times at which the instantaneous transition rate diverges into a two-dimensional hierarchy indexed by lattice winding numbers $(m, n)$. This hierarchy reflects the structure of the winding lattice $\Lambda$ and provides a direct probe of the torus geometry. The safe observation window is determined by the smallest critical time $s_c^{\min}$. It is generically shorter than in the single-compactification case at comparable length scales. This is because the two-dimensional lattice packs more short vectors than any one-dimensional sub-lattice. In principle, a sufficiently detailed measurement of the divergence pattern may encode enough information to reconstruct the winding lattice up to an overall rotation, thereby determining both $L_1$ and $L_2$.

For the detector accelerating in the noncompact direction, the topological correction \eqref{eq:topo_correction} multiplies both channels through the same bracket, so it cancels in the excitation to de-excitation ratio \eqref{eq:detailed_balance}. The detector satisfies detailed balance at the Unruh temperature $T = 1/(2\pi\alpha)$ regardless of $L_1$ and $L_2$: compactification alters the overall rate but not the temperature the detector reports. This separation is a statement about the pole ladder of the pulled-back Wightman function. The temperature is set by the vertical spacing $2\pi\alpha$, fixed by the acceleration alone, while the identifications move poles horizontally within a row.

A unifying observation across the three cases is the role of the lattice norm $\ell_{mn} = \sqrt{(mL_1)^2 + (nL_2)^2}$. In the inertial case it appears inside the relativistic distance $\Rmn$; in the compact-acceleration case it sets the critical times through the implicit equation \eqref{eq:27}; and in the noncompact-acceleration case it controls the argument of the topological correction \eqref{eq:topo_correction}. The dependence on $\ell_{mn}$ rather than on $mL_1$ and $nL_2$ separately is what makes the transition rate sensitive to the aspect ratio $L_1/L_2$ and not merely to the total area. This aspect-ratio sensitivity is a genuinely two-dimensional effect with no counterpart in the single-compactification analysis of \cite{Chiou:2016exd}.

Table~\ref{tab:limits} collects the main consistency checks satisfied by our results in the relevant limiting cases.

\begin{table}[h!]
\centering
\renewcommand{\arraystretch}{1.35}
\begin{tabular}{p{3.6cm}p{4.2cm}p{6.2cm}}
\hline
\textbf{Limit} & \textbf{Trajectory} & \textbf{Result recovered} \\
\hline
$L_1, L_2 \to \infty$ & inertial, Sec.~\ref{sec:inertial} &
$\Feq \to -\Delta E/(2\pi)$, eqs.~\eqref{eq:excitation_zero}--\eqref{eq:velocity_derate_real} \\
$L_1, L_2 \to \infty$ & noncompact accel., Sec.~\ref{sec:accel_noncompact} &
$\Corr \to 0$, pure Planckian spectrum in both channels, eq.~\eqref{eq:noncompact_rate} \\
$L_1 \to \infty$, $L_2$ fixed & all three &
all $m \neq 0$ terms drop; the single-compactification case, with the full pole ladder of
\eqref{eq:noncompact_poles} retained in Sec.\ref{sec:accel_noncompact} \\
$\alpha \to \infty$ & noncompact accel., Sec.~\ref{sec:accel_noncompact} &
inertial result \eqref{eq:velocity_derate_real} at $v_z = 0$ \\
$L_1, L_2 \gg \alpha$ & compact accel., Sec.~\ref{sec:accel_compact} &
only the $(0,0)$ image survives; stationarity and the Unruh--Planck spectrum \\
$\ell_{mn} \to 0$ & noncompact accel., Sec.~\ref{sec:accel_noncompact} &
image summand $\to \Delta E/(2\pi)$, matching the Rindler double pole \\
$\Feq(\Delta E)/\Feq(-\Delta E)$ & noncompact accel., Sec.~\ref{sec:accel_noncompact} &
$e^{-2\pi\alpha\Delta E}$ for all $L_1, L_2$; detailed balance at $T = 1/(2\pi\alpha)$ \\
\hline
\end{tabular}
\caption{Limiting cases and consistency checks. The identity $\Finst(-|\Delta E|) - \Finst(|\Delta E|) = |\Delta E|/(2\pi)$ of eq.~\eqref{eq:exc_deexc_relation} holds whenever $\mathcal{S}(\tau,s)$ is regular on $(0,\Delta)$, which is the case for the switched detector of Sec.~\ref{sec:accel_compact} below the first critical time.}
\label{tab:limits}
\end{table}

There are several natural directions for future work. On the inertial side, it would be useful to derive the full response for motion with nonzero velocity components along both compact directions, and more generally for trajectories not aligned with the coordinate axes. On the geometric side, one may ask whether a detector can distinguish a general flat two-torus, including its shape modulus, and not only the rectangular case studied here. It would also be interesting to extend the analysis to massive or fermionic fields, alternative boundary conditions, and multi-detector observables such as entanglement harvesting.

\appendix

\section{Contour Integration for Section~\ref{sec:accel_noncompact}}
\label{app:contour}

We present the residue computation that leads from the Wightman function \eqref{eq:noncompact_wightman} to the equilibrium transition rate \eqref{eq:noncompact_rate}. The starting point is the equilibrium rate formula \eqref{eq:eq_rate} applied to the pulled-back Wightman function,
$$\dot{F}_{\rm eq}(\Delta E) = \int_{-\infty}^{\infty} ds\; e^{-i\Delta E\,s}\; W_T(s),
\qquad
W_T(s) = -\frac{1}{16\pi^2\alpha^2}\sum_{m,n}\frac{1}{\sinh^2\left(\frac{s}{2\alpha} - \frac{i\epsilon}{2\alpha}\right) - \frac{\ell^2_{mn}}{4\alpha^2}}.$$

\noindent\textit{Pole ladder.} For each $(m,n)$ the denominator vanishes when $\sinh\bigl(\frac{s-i\epsilon}{2\alpha}\bigr) = \pm\frac{\ell_{mn}}{2\alpha}$. Since $\sinh(u + i\pi) = -\sinh u$, and only $\sinh^2$ appears, every solution may be shifted by any integer multiple of $i\pi$ in $u = (s-i\epsilon)/2\alpha$. The complete pole set is therefore \eqref{eq:noncompact_poles}: a pair of simple poles at horizontal positions $\pm 2\alpha\,\mathrm{arcsinh}(\ell_{mn}/2\alpha)$, repeated in every row at height $2\pi\alpha k + \varepsilon$. For $(m,n) = (0,0)$ the two positions coincide and the pole is second order. The $i\epsilon$ prescription places the $k=0$ row just above the real axis, so the rows at $k \geq 0$ lie in the upper half-plane and those at $k \leq -1$ in the lower half-plane.

\noindent\textit{Row-$0$ residues.} For the $(0,0)$ term, $\sinh(s/2\alpha) \to s/2\alpha$ and $4\alpha^2\sinh^2(s/2\alpha) \to s^2$ as $s \to 0$, while $e^{-i\Delta E s} = 1 - i\Delta E s + O(s^2)$. The coefficient of $1/s$ in $-\frac{1}{4\pi^2}s^{-2}(1 - i\Delta E s)$ is $i\Delta E/(4\pi^2)$. For $(m,n) \neq (0,0)$ write $f = N/D$ with $N = -\frac{1}{4\pi^2}e^{-i\Delta E s}$ and $D = 4\alpha^2\sinh^2(s/2\alpha) - \ell^2_{mn}$, so that $D' = 4\alpha\sinh(s/2\alpha)\cosh(s/2\alpha)$. Putting $\beta_{mn} = \mathrm{arcsinh}(\ell_{mn}/2\alpha)$ and using $\cosh\beta_{mn} = \sqrt{\ell^2_{mn}+4\alpha^2}/(2\alpha)$ gives $D'(\pm 2\alpha\beta_{mn}) = \pm\ell_{mn}\sqrt{\ell^2_{mn}+4\alpha^2}/\alpha$. Adding the two residues and using $e^{ix}-e^{-ix} = 2i\sin x$,
$$\mathrm{Res}^+ + \mathrm{Res}^- = \frac{i\,\alpha\,\sin\bigl(2\alpha\Delta E\,\beta_{mn}\bigr)}{2\pi^2\,\ell_{mn}\sqrt{\ell^2_{mn}+4\alpha^2}}.$$
Both row-$0$ residues are $i$ times a real number and both are odd in $\Delta E$. Multiplying by $-2\pi i$ gives the real quantity
$$-2\pi i\,r(\Delta E) = \frac{\Delta E}{2\pi} + \sum_{(m,n)\neq(0,0)}\frac{\alpha\,\sin\bigl(2\alpha\Delta E\,\beta_{mn}\bigr)}{\pi\,\ell_{mn}\sqrt{\ell^2_{mn}+4\alpha^2}} = \mathcal{B}(\Delta E),$$
which is \eqref{eq:row_zero}--\eqref{eq:topo_correction}, since $\sqrt{\ell^2_{mn}+4\alpha^2} = 2\alpha\sqrt{1+(\ell_{mn}/2\alpha)^2}$.

\noindent\textit{Summing the ladder.} Shifting row $0$ up by $2\pi i \alpha k$ leaves $W_T$ unchanged, because $\sinh^2$ has period $2\pi i\alpha$ in $s$, and multiplies $e^{-i\Delta E s}$ by $q^k$ with $q = e^{2\pi\alpha\Delta E}$. The residue of row $k$ is therefore $r\,q^{k}$.

For excitation ($\Delta E > 0$) the factor $|e^{-i\Delta E s}| = e^{\Delta E\,\mathrm{Im}\,s}$ decays downward, the contour is closed in the lower half-plane, and the enclosed rows are $k = -1, -2, \ldots$, traversed clockwise:
$$\dot{F}_{\rm eq}(\Delta E) = \bigl[-2\pi i\,r(\Delta E)\bigr]\sum_{k\geq 1} q^{-k} = \frac{\mathcal{B}(\Delta E)}{q-1}.$$
For de-excitation the exponential decays upward, the contour is closed in the upper half-plane, and the enclosed rows are $k = 0, 1, 2, \ldots$, traversed counterclockwise. Using $r(-\Delta E) = -r(\Delta E)$, the change of orientation and the change of sign of $r$ cancel, so the prefactor is again $\mathcal{B}(\Delta E)$:
$$\dot{F}_{\rm eq}(-\Delta E) = \bigl[+2\pi i\,r(-\Delta E)\bigr]\sum_{k\geq 0} q^{-k}\Big|_{q\to q^{-1}} = \frac{q\,\mathcal{B}(\Delta E)}{q-1}.$$
The two geometric series differ only in that the upward one starts at $k = 0$ and the downward one at $k = -1$, and a single extra term of a geometric series is exactly a factor $q$. This yields \eqref{eq:excitation_unruh}, and both cases are contained in \eqref{eq:noncompact_rate}. Dividing the two results gives \eqref{eq:detailed_balance}.

To close the contours one cannot use a large semicircle, since it would cut through the rows; the arcs are threaded between consecutive rows and then taken to infinity, where the summand decays exponentially in $|\mathrm{Re}\,s|$.

\noindent\textit{Numerical check.} Equation \eqref{eq:noncompact_rate} was checked against direct numerical evaluation of $\int ds\,e^{-i\Delta E s}W_T(s)$ along a contour in the pole-free strip $-2\pi\alpha < \mathrm{Im}\,s < 0$, for both signs of $\Delta E$ and for several values of $\alpha$, $L_1$ and $L_2$. The two agree to the precision of the quadrature, and the ratio \eqref{eq:detailed_balance} is reproduced to machine precision.

\section{ Numerical Methods}
\label{app:numerics}

We summarize the procedures used to generate Figures~\ref{fig:velo_fig}--\ref{fig: Figsect6}. The complete source code is available at \url{https://github.com/Sheeshram-siddh/Local-Probes-of-Global-Topology}.

\paragraph{Lattice truncation.}
All three sections involve sums over the winding lattice $(m,n) \in \mathbb{Z}^2 \setminus \{(0,0)\}$. A mode is included if its lattice norm satisfies $\ell_{mn} < \Lambda$, and the retained modes are summed in order of increasing $\ell_{mn}$. This isotropic cutoff avoids the direction-dependent truncation error of a rectangular box $|m|,|n| \leq N$, which reaches a distance $\sim N\sqrt{L_1^2+L_2^2}$ along the diagonals but only $NL_1$ or $NL_2$ along the axes. The values used are $\Lambda = 10\,000$ for Figure~\ref{fig:velo_fig}, $\Lambda = 500/\Delta E$ for Figure~\ref{fig:Figsection5} and $\Lambda = 1\,000/|\Delta E|$ for Figure~\ref{fig: Figsect6}, each checked by doubling $\Lambda$.

\paragraph{Sampling of Figures~\ref{fig:zoomL1} and \ref{fig:zoomL2}.}
These use the same expression and the same cutoff as Figure~\ref{fig:velo_fig}; only the plot window and the sampling step differ, as stated in the captions.

\paragraph{Section~\ref{sec:accel_compact}.}
The $(0,0)$ piece of $\mathcal{S}(\tau,s)$ behaves as $-1/s^2 + 1/(12\alpha^2) + O(s^2)$ near $s=0$. The counterterm cancels the leading divergence and the combination has the finite limit $1/(12\alpha^2) + \Delta E^2/2 + \sum_{(m,n)\neq(0,0)}\ell_{mn}^{-2}$, which we impose analytically for $s < 10^{-4}$ to avoid cancellation of two large numbers. The remaining integral is evaluated by adaptive Gaussian quadrature with tolerances $10^{-5}$, on a mesh that concentrates points near $\Delta \to s_c^{\min}$. The critical durations \eqref{eq:27}--\eqref{eq:critical} are solved by bisection; the smallest root over all $(m,n)$ sets the integration range, which is stopped at $\Delta = s_c^{\min} - 0.005/\Delta E$.

\paragraph{Section~\ref{sec:accel_noncompact}.}
The Planckian prefactor in \eqref{eq:noncompact_rate} overflows for $|2\pi\alpha\Delta E| > 500$ and is replaced there by its leading asymptotic form. Equation~\eqref{eq:noncompact_rate} was verified against direct numerical evaluation of the Fourier transform, as described in Appendix~\ref{app:contour}.

\paragraph{Decompactification proxy.}
Curves labelled ``$L = \infty$'' or ``Pure Unruh'' are computed by setting the relevant length to $10\,000/\Delta E$, large enough that the corresponding winding modes fall outside the cutoff.

\section*{Declaration of generative AI and AI-assisted technologies in the
manuscript preparation process}

During the preparation of this work, the authors used \textsf{Claude AI (Anthropic)}
in order to obtain assistance with language editing, to improve clarity,
readability and phrasing. After using this tool, the authors reviewed and
edited the content as needed and take full responsibility for the content of
the published article.


\bibliography{main}

\end{document}